# THE STORY OF TELEBRAIN:
## A multi-performer telematic platform for performatization

**Abbreviated Title: The Story of Telebrain**

**Kristin Grace Erickson**

Performer-Composer D.M.A. Program
The Herb Alpert School of Music
California Institute of the Arts
24700 McBean Pkwy
Valencia, CA 91355
United States of America
kerickson@calarts.edu
+1 (850) 656-8879

# ABSTRACT

This paper presents Telebrain, a browser-based *performatization* platform invented for organizing real-time telematic performances.

Performatization is the human performance of algorithms. When computers and humans performatize cooperatively, the human-computer interaction (HCI) becomes the location of computation. Novel modes of machine-human communication are necessary for organizing performatizations. Telebrain is designed to facilitate machine-human languages.

Capitalizing on the ubiquity and cross-platform compatibility of the Internet, Telebrain is an open-source web application supporting PerPL (Performer Programming Language), a human-interpreted configurable language of multi-media instructions used to program performers.

Telebrain facilitates a variety of performance disciplines such as music, theater, dance, computational performance, networked scoring (image and audio), prompted improvisation, real-space multi-player gaming, collaborative transdisciplinary karaoke and quantum square-dancing. (http://telebrain.org)

# INTRODUCTION

Visualization, sonification, and auralization are established practices for representing data and computational outputs. This paper develops *performatization* as the practice of performing data, and of performing computation on data, through music, theatre, dance, improvisation, and other transdisciplinary frameworks. A multi-performer, real-time, browser-based telematic platform for distributing a large volume of performatization cues is introduced for creating large scale human-performed computations through the arts. This telematic performatization platform is called Telebrain.

My recent work has involved the iterative design of performatization systems, entangling research and development with a collaborative artistic practice. This has resulted in a non-linear history of complex intersections between performance and computation. Heuristic methods of performatization originated from the analysis of improvised music and theater games. Both used rules to organize improvisations facilitating simple translations between performance instructions and computer algorithms. Creative experiments distributing performance instructions culminated with the invention of Telebrain.

Telebrain is a web-based performatization platform for collectively developing and distributing PerPL (Performer Programming Language), pronounced *purple.* A multi-media, multimodal programming language for machine-human communication, PerPL integrates language, logic, and art into a single symbolic instructional paradigm. Allowing evolutionary specification, PerPL relies on syntax and context plasticity provided by human interpretation, adapting its rules through implementation.

PerPL programs are typically distributed by a computer to parallel human brains. PerPL

executes traditional SIMD (Single Instruction, Multiple Data) and MIMD (Multiple Intruction, Multiple Data) processing when multiple performers interpret the instructions — instructions are interpreted by *vectors* of performers.

By expanding computational architecture to include performance, the human-computer interaction (HCI) becomes a location for computation. As PerPL programs are interpreted by performers, computation emerges as artistic expression. Revolutionalized modes of communication, between creator, process, output, and observer, unleashes novel paradigms of creative potential yet to be fully conceived.

Although any type of performance can performatize, music functions doubly as performatization and as a mode of instruction. Music's ability to signify, encapsulate, sequence, and coordinate simultaneity contributes to a computational paradigm where perceivable machine transmissions are required.

In addition to serving as the launching point for conceiving performatization, sonification and auralization provide a scientific basis for applying the semiotic agency of music and sound to an audio-based human-interpreted programming language.

# BACKGROUND

**Sonification, Auralization and Performatization**

Sonification is the mapping of data into the audio domain. For thousands of years sound has been used to represent information heard as striking clocks and bell towers. Pythagoras used measurements of the stars and planets to define a musical scale. The stethoscope, geiger counter, and telegraph are all sonification inventions.[1]

Since the birth of Information Theory, the capacity for sound to transmit information has been researched to greater depths. Perceptual thresholds have been tested to determine the amount and types of multivariate data discernible through sound.[2] Some sonification techniques encode data into the acoustic parameters of sound waves while other techniques map data into culturally-determined systems of musical organization.[3]

Auralizations, also called Auditory Displays, are a special type of sonification where sound represents the output of a computer process.[4] Considered the audio equivalent of a computer monitor, auralization is used to communicate hierarchical and navigational information to blind computer users.[5]

Auralization research has also focused on warning signals in airplane cockpits, since pilots are inundated with information and certain data needs to make itself known in high-stress conditions.[6] Different types of warning signals have been developed and tested for effectiveness in relation to the types of data needing to be communicated and the cognitive response to the sounds.[7] The *performance* of piloting a plane is aurally-informed by audio representations of the plane's internal systems in addition to the atmospheric systems through which the plane is flown.

Extending sonification and auralization, performatization is the human performance of algorithms. Unlike visual displays and auralizations, where pixels and sounds illustrate the output of a computer process, performatizations *are* the algorithmic process. Performatizations compute.

**Performatizing Bubble Sort**

The following description of a hypothetical performatization refers to a sorting algorithm found in most introductory algorithm textbooks.

The Bubble Sort algorithm begins by comparing the first two values in a list. If the values are out of order, Bubble Sort swaps the positions of the two values before comparing the next two values. When the end of the list is reached, Bubble Sort performs another iteration by returning to the beginning of the list and repeating the comparison process. If no further swapping is necessary at the end of an iteration, the sort is complete. Performatizing a Bubble Sort algorithm involves performing the individual comparisons and iterations of the algorithm.

To prepare, each performer wears a different number on their shirt. The performance begins by randomly shuffling a group of performers into a single line — into an array. An additional performer named *Deictor* functions as an array pointer and binary flag. *Deictor* begins each iteration by literally pointing to the first performer in the line. If the number worn by the performer to the left is less than the number worn by the performer, the two performers swap positions in line. *Deictor* steps through the list by pointing to the next performer in line and performing additional comparisons.

To indicate when the sort is complete, *Deictor* raises one arm at the beginning of each iteration, lowering the arm when a swap occurs. The raised arm functions as a binary flag signaling if additional iterations are required. If *Deictor's* arm is down at the end of an iteration, the list is not sorted and requires another iteration. If *Deictor's* arm is raised at the end of an iteration, the sort is complete and the performers are in numerical order from left to right.

This example can be extended to singing. Instead of numbers worn on clothing, each performer is assigned a unique pitch to sing. The performers sing their notes to compare pitches. If the pitches are out of order, the performers swap positions in line. Each iteration through the line produces a unique melodic contour and the sorting process is heard as melodic lines gradually developing toward an ascending pattern of notes.

Generally, computer science is concerned with algorithm efficiency, linking the qualitative value of an algorithm to the amount of resources a computational process requires. When performatizing, efficiency is overridden by the aesthetic quality and experiential value of the performance. In experimental performance, breaking rules reveals novel performance potential — malfunction leads to success, to change. Relieving algorithms of their need to achieve may expand the cultural applications of computation.

In Bubble Sort, the predetermined outcome is the *sorted* state of the list, and the algorithmic process produces this desired outcome. In traditional performance, the predetermined outcome is the performance itself. Performatizations regard performances as processual, and therefore a location where algorithms compute, as shown in Image 1.

**Image 1. Architecture Comparisons**

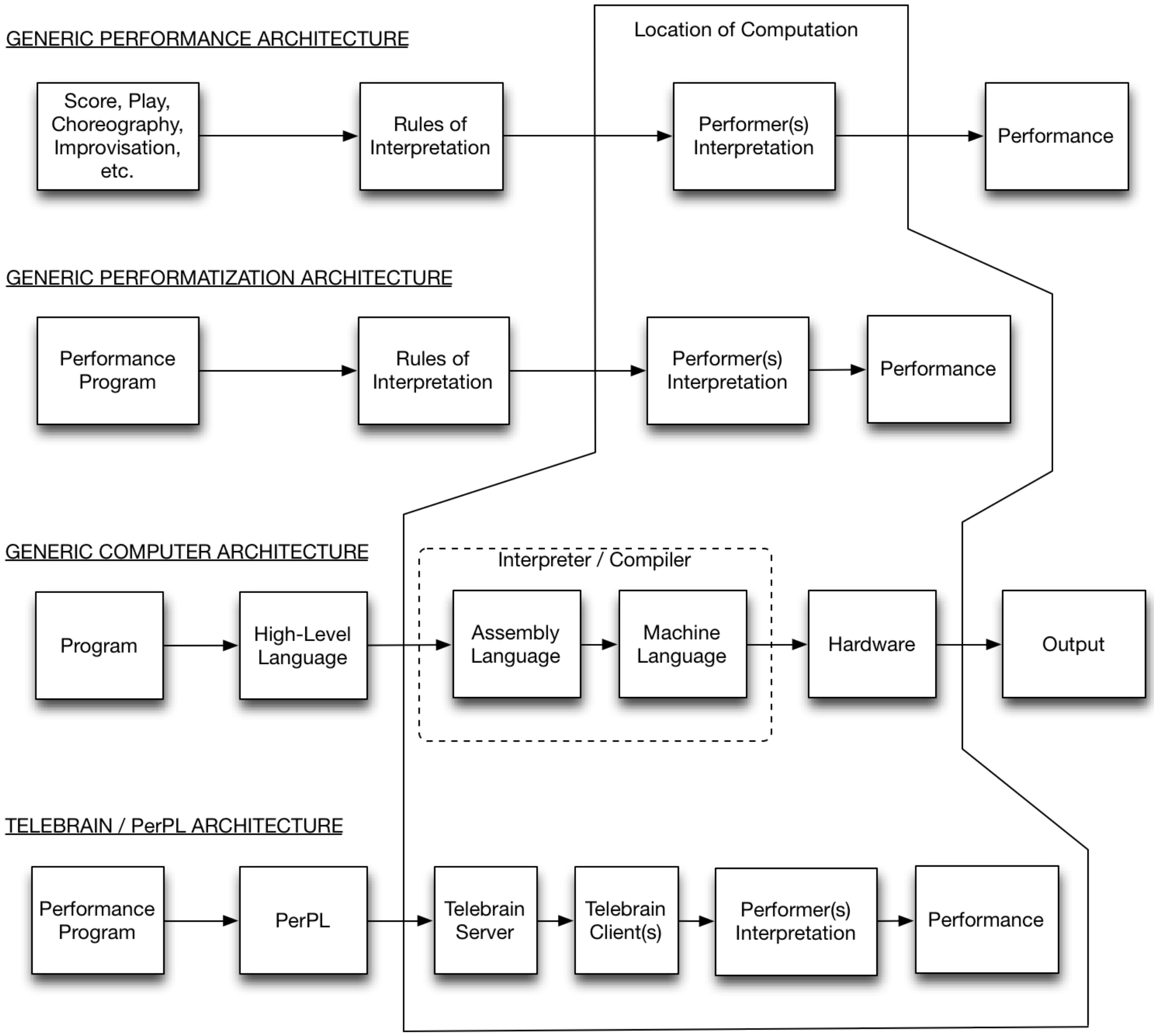

Removing the goal-state of the Bubble Sort performatization undermines the necessity for accurately evaluated comparisons. Now, an individual comparison can result in two players swapping positions because they want to. Performers have brains. Introducing human decision-making into an algorithmic process results in an algorithmic process capable of self-determining its outcome.

From an aesthetic standpoint, an ascending pattern of notes is not necessarily the preferred outcome of a Bubble Sort performatization. The experience of singing and listening to the melodic contours of previous iterations informs the swapping decisions of subsequent iterations, systematically democratizing aesthetic decisions of collaboration. Boolean probabilities correlate to applications of aesthetic criteria. Logic gates built from human personality and interaction self-organize patterns of melodic contour iterations, entangling computation, game theory, and aesthetics.

Rules of performatizations must be communicated to and understood by performers. Everyday language can explain simple algorithms, but performatizing complex processes requires a configurable language optimized for communicating real-time processes to multiple performers.

**Cobra**

> … the pieces slowly evolved into complex on-and-off systems … I eliminated the timeline. What remained were scores that did not refer to sound or time — two parameters traditionally inseparable from the art of music — but were a complex set of rules that, in a sense, turned players on and off like toggle switches to such a complicated degree that it didn't really matter what the content was.
>
> — John Zorn, "The Game Pieces"

John Zorn's improvised music game *Cobra* serves as a key model for *performing a system* using a configurable metalanguage to organize performances. The rules of *Cobra* enable self-organizing performance interactions to collectively embody a complex configurable system.

*Cobra* is played using color-coded cue cards inscribed with symbols. Players use hand signals to indicate desired cues to a prompter. The prompter presents the requested cue card to the ensemble before activating the cue by lowering the card. Cues are often called in quick succession causing sudden shifts in improvised material to create a fragmented musical experience. *Cobra* includes an embedded game allowing guerrilla factions to subvert the prompter/player relationship and subsequent music.

Cues initiate new musical information or modify existing musical information. Some cues require everyone to play while others require specific player assignments or sequence. Memory Cues associate played music with placeholders so previous musical states of the game can be recalled.

*Cobra* illustrates an efficient and practical communication system for organizing music improvisation, but not without limitations. Since *Cobra* relies on musical material improvised by experienced musicians, the cues designate imprecise qualities of music bypassing deeper levels of specificity. A performatization language designed to maximize its potential applications necessitates a capacity to communicate between extremities of precision. Since cue cards require continual eye contact between player and prompter, the system is impractical for performances involving body movement. To resolve the performance constraints of *Cobra's* visual cueing system, audio cueing systems were investigated for communicating further gradations of instruction specificity.

## M.T.BRAIN

M.T.Brain (Music Theater Brain) is a Max/MSP patch developed for distributing real-time audio instructions to multiple performers. Using an audio routing matrix, parallel channels of audio are sent to performers through a ten-channel sound card, long cables, and headphones. Performers wear large numbers because performer interactions are indicated by channel number, as shown in Image 2.

**Image 2. M.T.Brain Performance**

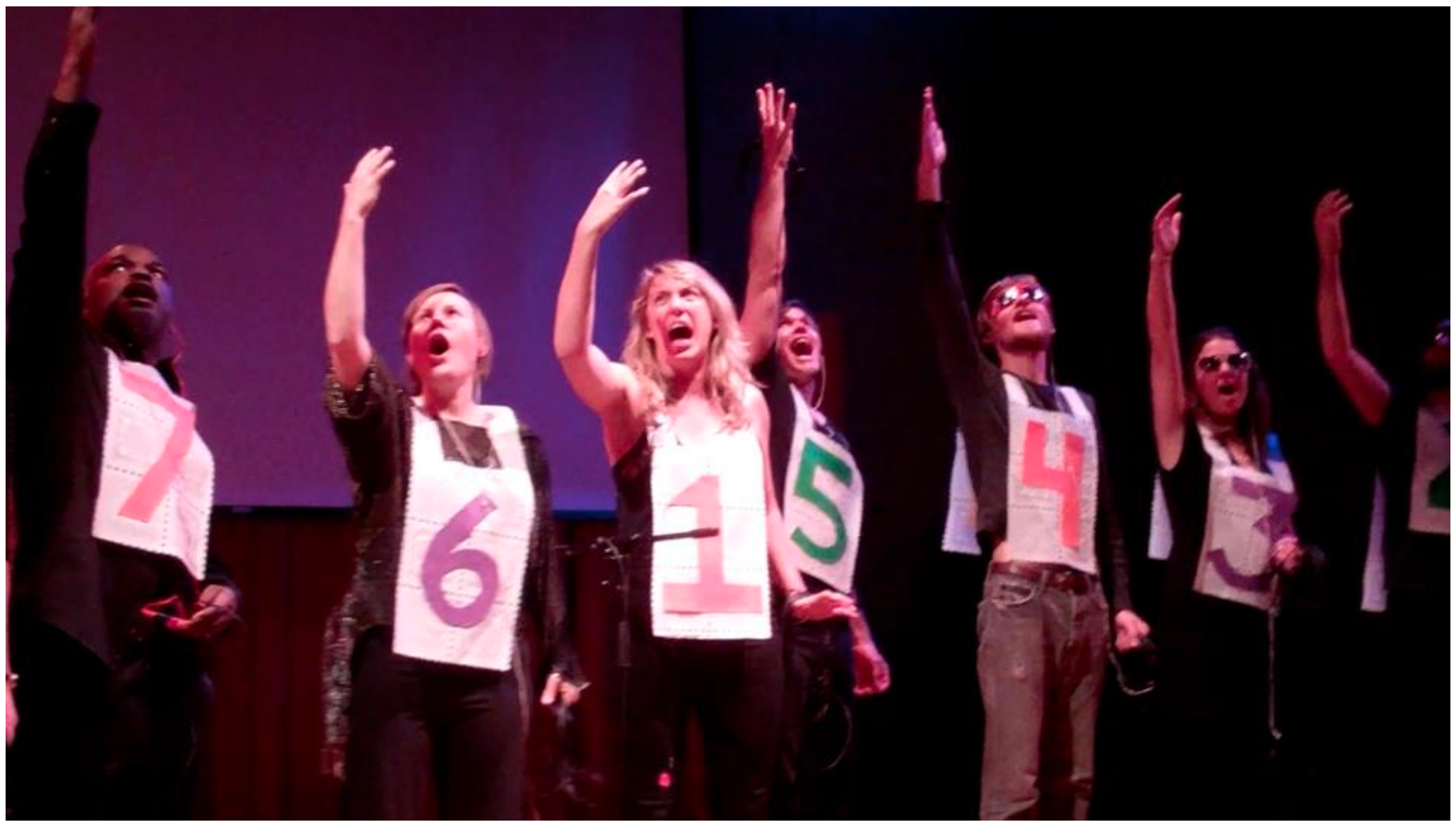

A prompter operating M.T.Brain routes preset and real-time audio instructions to performers, as shown in Image 3. Typed instructions are spoken by an artificial text-to-speech voice. During a performance, new instructions can be saved for repeated use. A collection of easily-identifiable non-speech sounds are used to represent learned instructions or to indicate precisely when to perform an instruction. Like components of a configurable symbolic language, audio cues are concatenated with text-to-speech outputs to construct longer instructions, a prototype of PerPL.

**Image 3. M.T.Brain Max/MSP Prompter Interface**

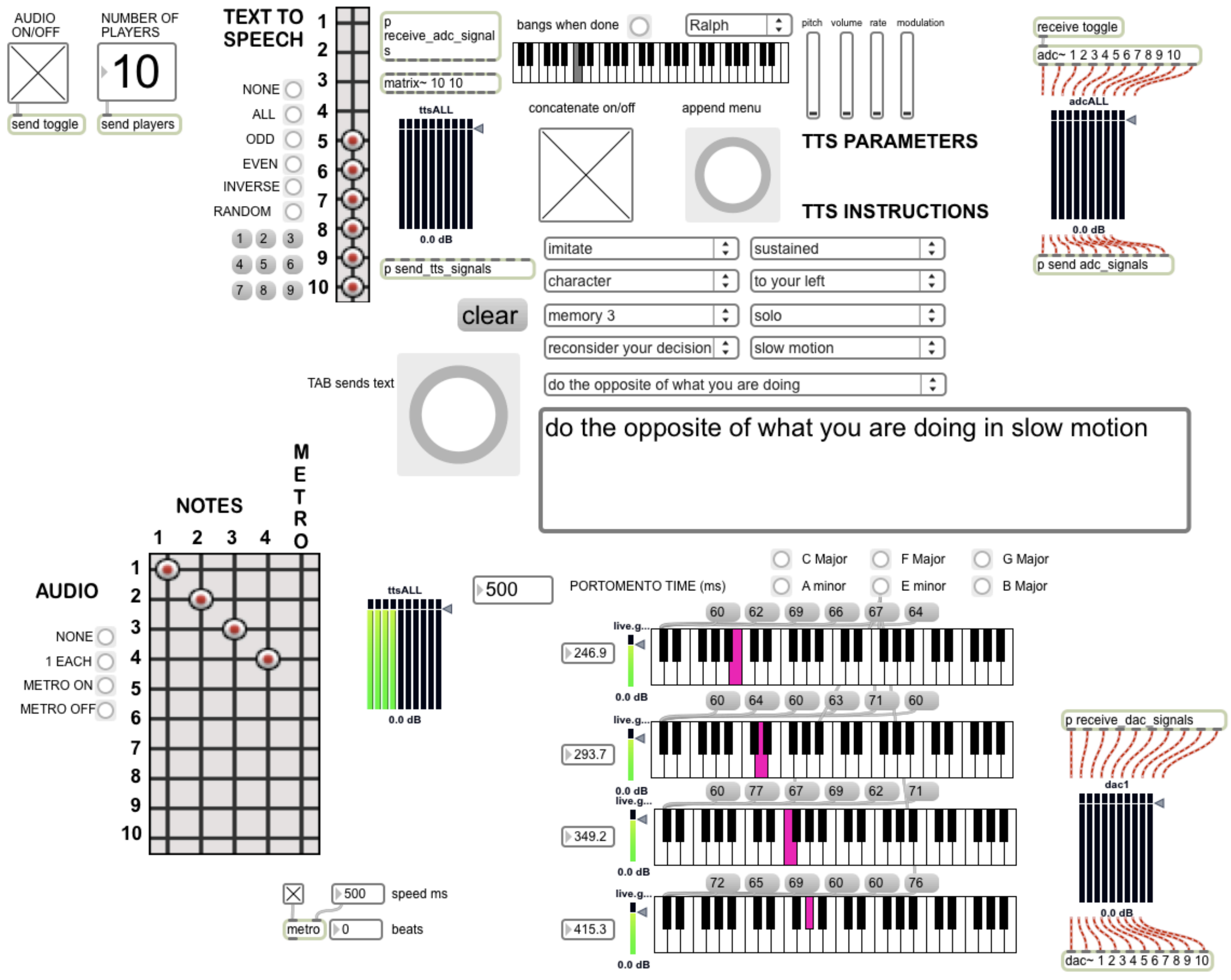

Adjusting to the number of performers, M.T.Brain provides shortcuts for performer routing configurations including: all channels, one channel, odd or even channels, randomly selected groups of channels, and unselected complementary channels.

A metronome and four sine tones with frequency presets can be routed concurrently with audio instructions. Ten microphone inputs can be routed through M.T.Brain enabling real-time spoken instructions to be heard during a performance, as shown in Image 4.

**Image 4. M.T.Brain Architecture**

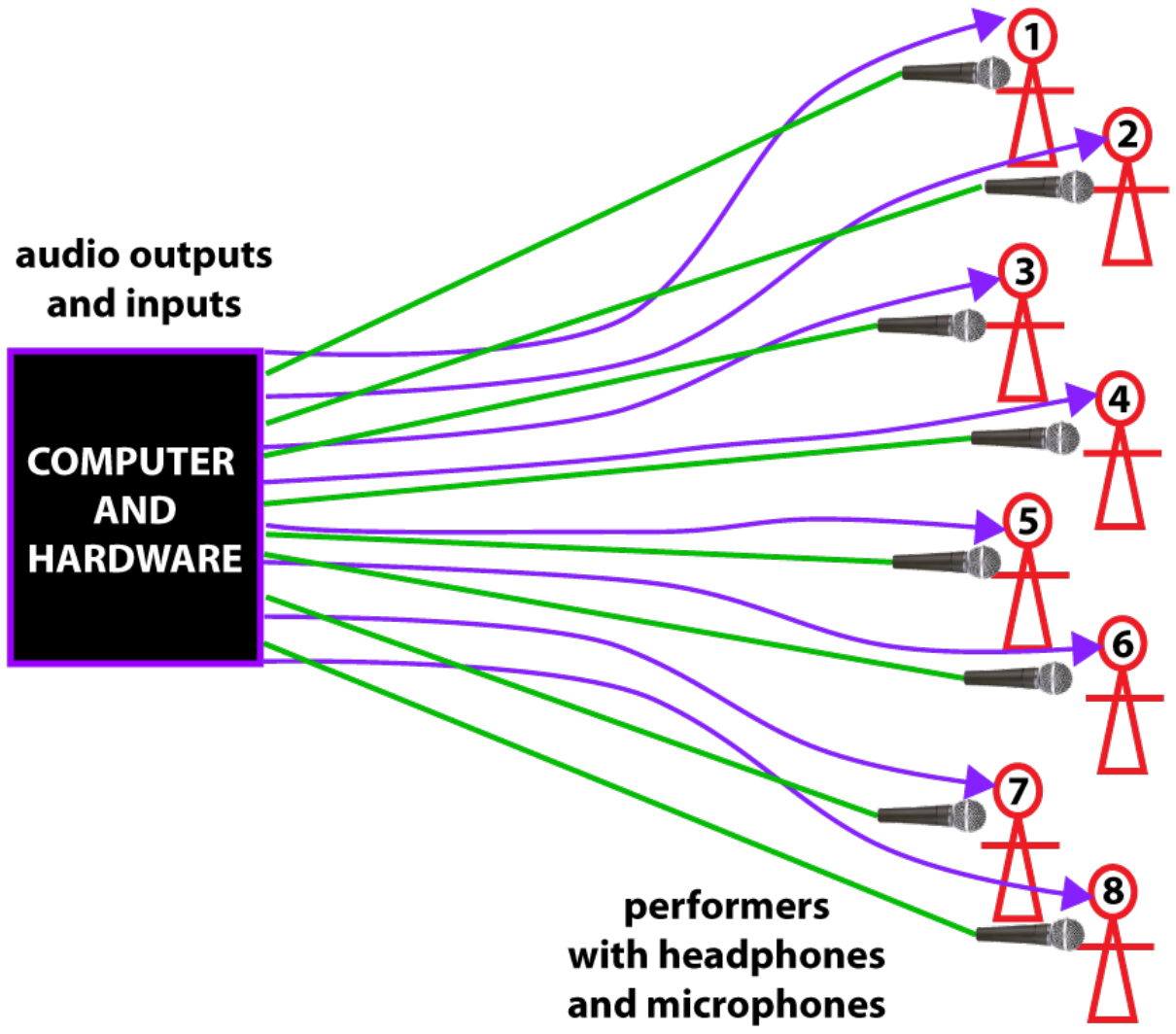

Preset audio cues inspired by improvised music and theater games can be used to organize live collaborative transdisciplinary improvisation. Styles of M.T.Brain performances are influenced by the prompter's approach to cueing and routing. Prompters can extend traditional modes of performance organization by configuring M.T.Brain to support the desired outcome.

For musical results, a prompter can take advantage of the temporal accuracy resulting from M.T.Brain's hard-wired parallel audio connections. By algorithmically scattering time-dependent instructions between multiple channels, so each performer receives 1/10th of the instructions constituting a complete gesture, the immediate link from computer-to-audio-to-ear-to-brain gives rise to tight rhythmic precision and multi-performer gestures to be heard — with little or no rehearsal. M.T.Brain performers can function like hammers on a player piano.

A prompter can provoke a spontaneous song and dance routine by routing lyrics to a lead vocalist who improvises the melody of a song. Simultaneously, the four sine tones are routed to the remaining performers who sing an accompaniment based on the four-voice chord presets the prompter triggers. M.T.Brain's parallel in-ear sine tones simplify singing dissonant and microtonal harmonies and changes. The sine tones are layered with a metronome and audio instructions for synchronizing choreography instructions distributed according to channel number.

For theatrical performance, the dialogue and stage directions of a play, prewritten or prompter-improvised, are individually routed to each character. One of the audio channels is used to cue a lighting engineer while two other audio channels are routed to an amplification system for stereo sound effects and incidental music to be heard, timed with theatrical action.

Despite an abundance of potential applications spanning disciplines, the long cables connecting the sound card to the performers' headphones are a serious physical obstacle. As performers move through the space, they constantly wrangle their cables to minimize tripping hazards. The cables tangle with every movement into more complicated knots. M.T.Brain's overall aesthetic emerged as an web of wires ensnarling performers wearing large numbers.

**M.T.Brain iOS**

The M.T.Brain iOS app replaced cumbersome cables with wireless communication. The Max/MSP patch routes OSC (Open Sound Control) messages through a local wireless network to trigger text-to-speech audio and prerecorded audio files stored in the iOS application, as shown in Image 5.

**Image 5. M.T.Brain iOS Max/MSP Interface**

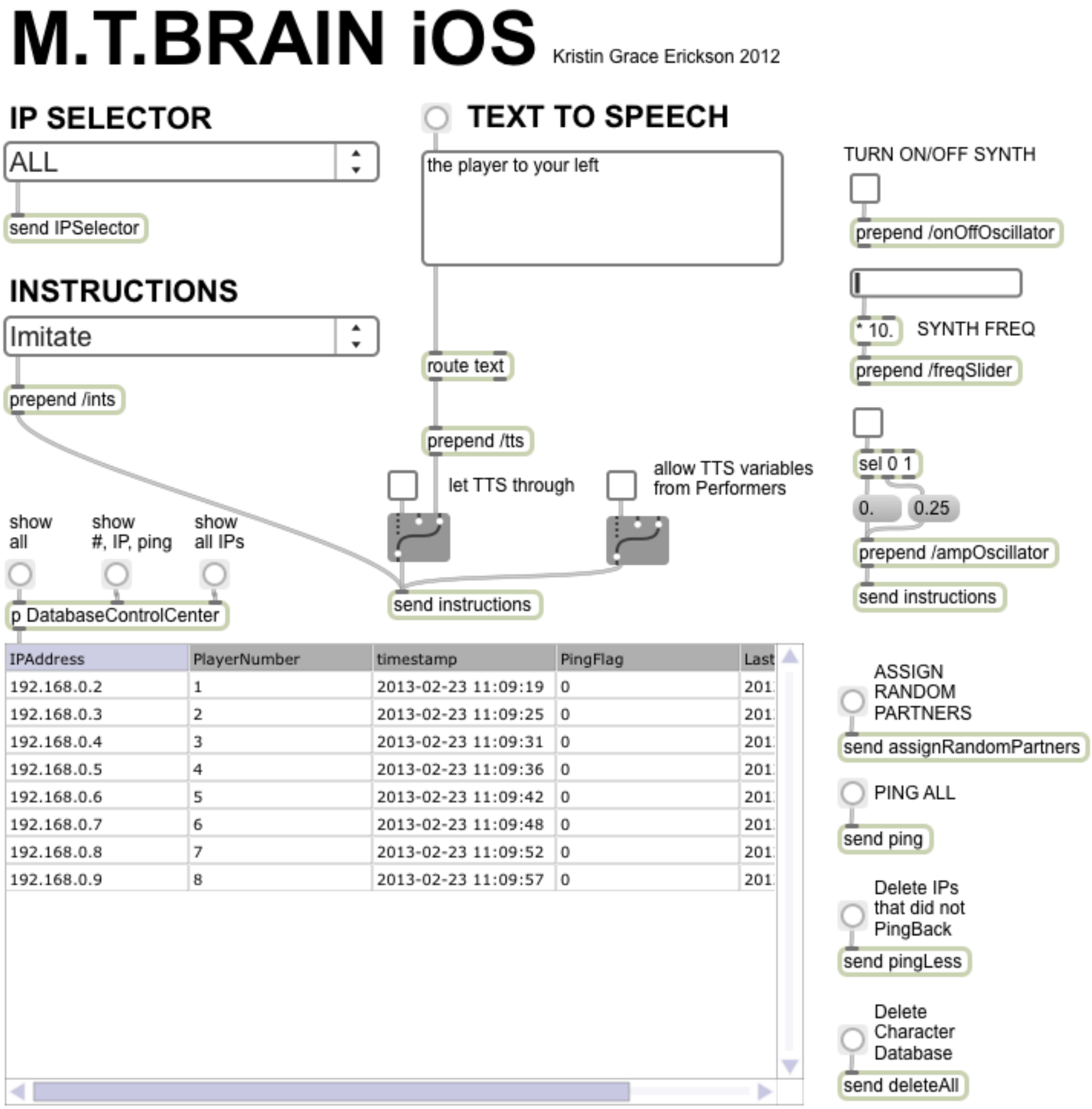

The iOS app includes an AM synthesizer to support instructions requiring pitch information. Audio is heard through the device's built-in speakers or headphone output.

Although wirelessly distributing OSC data through a network introduced inconsistent latency limiting the rhythmic accuracy available in the hard-wired M.T.Brain, the iOS app expanded existing M.T.Brain functionality with a user interface allowing each performer to contribute instructions during a performance, as shown in Image 6. The updated Max/MSP brain can run as a background OSC server while M.T.Brain performers prompt each other.

**Image 6. M.T.Brain iOS App Interface**

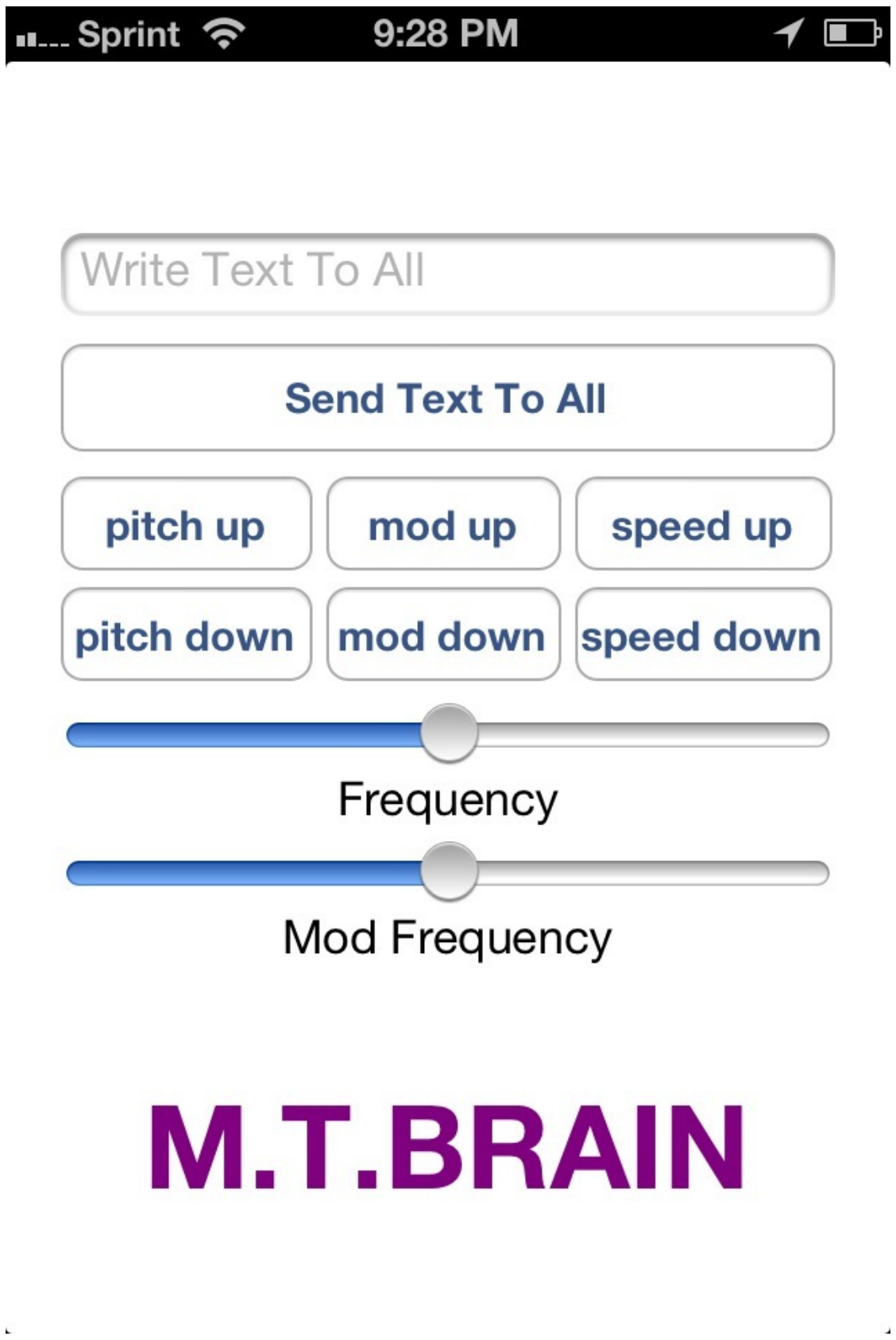

M.T.Brain iOS performances explore new modes of collaborative organization brought to light by performers participating in the prompting process. Decentralizing the role of the prompter replaced the aesthetic of wrangling cables with the equally pervasive aesthetic of staring at glowing phones. Nonetheless, I wrote two pieces using the M.T.Brain iOS app for audience members to perform.

In the first piece *Icebreaker,* the Max/MSP patch randomly assigns each phone's OSC output to another phone's OSC input by associating the local IP addresses. Each performer controls the frequency and modulation rate of an AM synthesizer on a randomly assigned phone. The participants walk around listening for the phone they control by sending contours recognizable over the hum of multiple modulated sine tones. When most of the participants have identified the phones they control, the Max/MSP patch is manually triggered to reassign new partners and begin again — telematic do-si-do.

The second piece *Garbledygook* utilizes the send-to-all text-to-speech function written into the iOS app. Standing on the edges of the room, each participant types a message, one at a time around the room. The text-to-speech artificial voice interpreting the message is heard on the built-in speakers of all of the phones with varying amounts of latency. After sending a message, each player takes one step towards the center of the room.

In multiple performances of *Garbledygook,* similar patterns of performance emerged. Performers type real language at first, inevitably discovering that random patterns of letters and differing amounts of letter repetitions yields unexpected audio results from the iPhone text-to-speech utility. When participants are surprised by new sounds, the messages that follow emulate and extend the discovered technique. The communication becomes more garbled and alien with every step towards the center of the room. By the end, the participants are huddled together, typing as fast as they can, giggling while their phones semi-simultaneously spit out a strange and new collective language.

Apple's licensing and distribution restrictions complicate the development and testing of applications intended for creative telematic collaboration. To bypass the limitations of developing for proprietary hardware, M.T.Brain was rewritten in JavaScript as an open-source browser-based application called Telebrain.

# TELEBRAIN

Telebrain is an internet performatization platform for creating and distributing real-time performance instructions to multiple performers using wireless web-enabled mobile devices and computers. ( http://telebrain.org )

Telebrain implements the performer programming language PerPL. Dependent on Telebrain functionality, PerPL is programmed using multi-media Content Objects, performer Role Objects, performance Venue Objects, Collections of stored associations, Multi-Role and Fractional instruction Assignments, Interface Objects, OSC Objects, timing Algorithms and instruction distribution Algorithms. The collective development of PerPL is facilitated online through Telebrain, where programmers store, organize, share and implement PerPL elements and instructions.

A complete description of Telebrain functionality is found in Appendix I. The priorities of Telebrain design are as follows:

- Ensure cross-platform and cross-browser compatibility
- Maintain a robust and extensible real-time performance infrastructure
- Create a platform for the collective development of PerPL
- Make PerPL instructions available to all users
- Protect PerPL elements and instructions from alteration or deletion
- Keep Telebrain design open to unforeseen applications and uses

PerPL instructions are distributed inside of instantiated Performance Venues that function like multi-media chatrooms. Instantiated Performance Venues enable complex patterns of PerPL instructions, represented as multi-media Content, to be distributed to designated performers in the Venue. PerPL is interpreted when performers receive and perform the instructions. Educating performers about PerPL syntax, semantics, grammar and data is the process of building a PerPL interpreter. Interpreter design evolves through iterative implementations of shared PerPL instructions.

**Telebrain Functionality**

Telebrain Content Objects include audio files, text-to-speech audio, images, and visually formatted texts. Telebrain programmers link to online media, upload media from local devices, or manually enter texts. Content Objects saved to Telebrain are organized into longer instructions using Collections, Associations, Assignments and Algorithms.

Content Objects are used to represent programming language elements such as variables, objects, operators, expressions, functions, assignments, conditionals, statements, and scope. Content Objects can also function as data, signifying preexistent meaning(s) of sounds, images, and words. An audio recording of a cat's meow can represent a variable or operator in PerPL, or the same audio file can be used as data representing a cat, a meow or a frequency contour.

Telebrain Collections organize associations and constraints. Folders group unordered content of any type. Audio-Image Pairs link Audio Objects with Image Objects simplifying distributions to multiple receivers with varying receive settings, or simultaneously distributing Audio and Image Objects to individual receivers. Audio Sentences are sequences of Audio Objects concatenated and saved as new audio files. The start and end times of the Audio Objects in an Audio Sentence are used to reorder Audio Sentences after being served to client devices. Image Phrases are sequences of Image Objects, incremented manually or by timing Algorithms during a Performance.

Role Objects determine the functionality of a performer's mobile device limiting the types of information performers can send and receive. Interface Objects are buttons, menus, and inputs associated with Objects, Collections, and Algorithms for customizing the layout and functionality of a performer's user interface.

Assignments simultaneously distribute unique instructions to multiple performers, an implementation of MIMD processing. Multi-Role Assignments associate PerPL instructions with specified Roles. Fractional Assignments associate PerPL instructions with fractional subgroups of performers.

Telebrain's default timing delivers Content as soon as possible, but the time required for Content to arrive on wireless clients cannot be accurately predicted. For messages to arrive on multiple wireless devices at the same time, a short delay is added to all instructions accommodating for differing speeds of delivery. Content received on client devices is held until a particular tick of the internal clock. Since the client clocks can be synchronized to Telebrain's server clock, previously served instructions can be executed simultaneously after compensating for network delays.

Telebrain Algorithms associate Timers, Metronomes, and programming logic with Content, Collections, Roles, and Assignments. The Timers and Metronomes control when images are displayed and audio is played. Algorithms allow OSC addresses to be associated with Telebrain functionality. External software can contribute to Telebrain Performances and external devices can receive OSC messages in response to Telebrain Performance activity.

Performing on Telebrain requires Roles and a Venue to be defined in advance. The first performer instantiates and names a Performance Venue, selects a Role, enters a nickname, and begins the Performance. Subsequent performers join the instantiated Performance Venue to participate in the Performance. Depending on Role definitions and associated Interface Objects, a Performance interface is rendered featuring lists of Content, Collections, Associations, and Algorithms available for execution. In the Performer List, checkboxes appear next to the names of performers and Roles. Instructions are routed according to Performer List designations. The Activity Log lists a record of timestamped performance instructions distributed during a Performance.

Performers in an instantiated Performance Venue can be in the same physical space as other performers or connected remotely through the Internet. Performances in multiple physical locations can occur within the same instantiated Performance Venue. Performers can leave and return to instantiated Performance Venues, but when the last performer leaves, the instantiated Performance Venue is destroyed.

As shown in Image 7, a screenshot of an active performer interface indicates the Performance is called *Free-For-All*, the performer's nickname is *Nick,* and Nick is performing a *Receiver* Role. Nick has sent the *F#4* Image Object to ALL performers including himself. A performer named *Bruno* is playing a *Lead* Role and Bruno previously sent Audio and Image Content to Nick.

**Image 7. Telebrain Performance Interface Example**

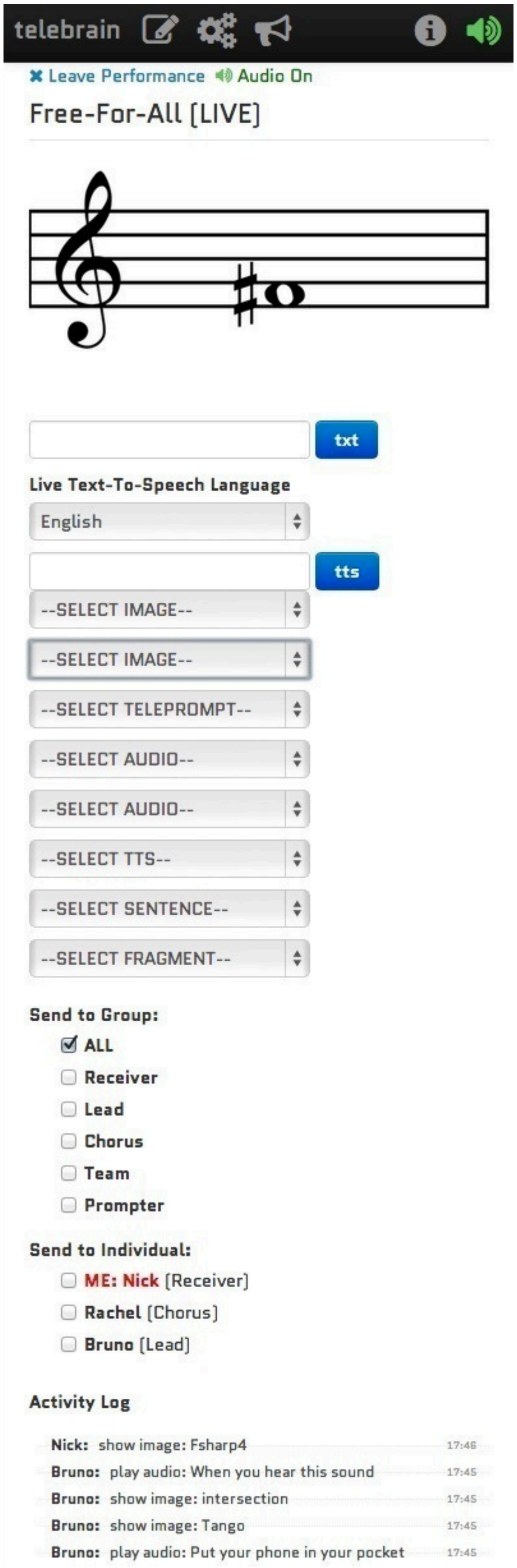

In addition to device memory, Telebrain performers function as memory. Like the Memory Cues in John Zorn's *Cobra,* Telebrain instructions can assign the current state of a performance to a Content Object, recalled by triggering the Content Object. Performer memory utilizes human memory capabilities and can be extended using external memory, such as writing information down on a piece of paper. Performers carrying a calculator or abacus can facilitate performatizations requiring accurately computed results from individual performer interactions.

**PerPL (Performer Programming Language)**

PerPL and Telebrain are designed to maximize configurability — operative as both specialized tools for specific applications and pliant tools for open-ended applications. Homologous with the Integrated Development Environments (IDEs) and language-aware text editors commonly used in computer programming, Telebrain governs the ways instructions are input, organized and delivered to PerPL interpreters. For each performance, protocols of PerPL syntax and semantics are agreed upon by PerPL programmers and Telebrain performers. PerPL's freely assignable syntax and semantics are informally stipulated through heuristic design, formed through Telebrain's infrastructure and performer capacity.

Explaining the execution of PerPL instructions to performers is the process of building multiple parallel PerPL interpreters. Training routines programmed in PerPL elucidate intended performance behavior and establish the meaning of shorthand instructions. Training can happen during warm-up routines prior to performing or training portions can be scattered throughout a Performance, introducing new protocols or shifting current rules mid-performance.

In Telebrain's Performance architecture, programmers can be performers, and vice versa.

Performances including instructions written by multiple programmers may introduce conflicting rules of syntax. When instructions are sent by the performers, for the performers, each training routine and its related instructions must indicate an association.

PerPL programmers must reinvent programming conventions to optimize human interpretation according to implementation. Techniques for indicating syntax, semantics, parsing, and scheduling require explicit definition and explanation for every Performance. Hidden, low-level programming conventions must be brought to the surface to enable high-level instructions to be performed.

In computer programming, quotation marks are often used to delimit strings. In PerPL, when devising string-delimiting techniques, the technique definition and an explanation of its function must be interpreted by the performer prior to its first use. In the following example, opening and closing quotations marks are assigned to Audio Objects. The actual quotation marks used in the example indicate phrases spoken by a text-to-speech voice.

```
//Quotation Mark PerPL Pseudocode

QuotationTraining = "When you hear this sound" + <<BEEP>> +
                    "vocally imitate what you hear next" +
                    "When you hear this sound" + <<BLEEP-BLOOP>> +
                    "stop imitating";

QuotationPractice = "Walk to the middle of the room" +
                    <<BEEP>> + "Jump up and down" + <<BLEEP-BLOOP>> +
                    "Look left" +
                    <<BEEP>> + "Now I will do it" + <<BLEEP-BLOOP>> +
                    "Jump up and down" +
                    <<BEEP>> +
                      <<(Sung)I am jumping up and down>> + "I will stop" +
                    <<BLEEP-BLOOP>> +
                    "Stop jumping" + "Look right" + "Walk out the door";
```

The sounds <<BEEP>> and <<BLEEP-BLOOP>> indicate if a performer jumps up and down, says the words "Jump up and down," or jumps up and down while singing the words "I

am jumping up and down."

Performers have varying response times to different types of instructions. Some performers act as soon as they begin to understand an instruction, while other performers wait for the entire instruction before acting. By incorporating sound-triggers into instructions, performer actions can synchronize.

Two-part sound-triggers provide preparatory timing clues for imminent actions. As an audio equivalent of seeing a conductor raise her baton before a downbeat, two-part sound-triggers contain audible *preparation* and *ictus* conducting gestures. To implement `Two-part Sound-Trigger Training`, the programmer uploads a two-part sound `<<shaa-CHUNK>>` into an Audio Object and then uploads the second half of the two-part sound `<<CHUNK>>` into another Audio Object. `<<CHUNK>>` is only used during the training routine to illustrate the sound of the *ictus,* the downbeat of the two-part sound-trigger.

```
//Two-Part Sound-Trigger Training

SoundTriggerTraining = "When you hear this sound" + <<shaa-CHUNK>> +
                       "Clap your hands at the exact moment you hear the" +
                       <<CHUNK>> + "portion of the sound";

//Two-Part Sound-Trigger Practice

      <<shaa-CHUNK>> [pause: 1000ms]
      <<shaa-CHUNK>> [pause: 8000ms]
      <<shaa-CHUNK>> [pause: 600ms]
      <<shaa-CHUNK>> [pause: 400ms]
      <<shaa-CHUNK>> [pause: 200ms]
      <<shaa-CHUNK>> [pause: 100ms]
      <<shaa-CHUNK>> [pause: 50ms]
      <<shaa-CHUNK>>
      <<shaa-CHUNK>>
      <<shaa-CHUNK>>
```

PerPL Audio Objects can function as continuous data controllers. As specifiable characteristics of a sound change through time, the contours can be correlated to the parameters of a performance. To illustrate, seven tap dancers receiving audio from Telebrain are instructed to change the speed of their tapping according to the fluctuating frequency of a sine tone instruction. In the following PerPL pseudocode, `<<(sine)__/~~~\_/\/~~~~~~\____/\/\/\/\/~~~~~~~~~>>` is used to represent a sine tone with a fluctuating frequency contour. The underscores, slashes, tildes and backslashes represent low, rising, high and falling frequency, respectively.

```
//Frequency-to-Tap-Dance Training

FreqTapTraining = "When you hear a sine tone" + <<(sine)>> +
                  "listen to its frequency" +
                  "When the frequency is high, tap fast" +
                  "When the frequency is low, tap slow" +
                  "As the frequency fluctuates between high and low
                  adjust your dancing speed according to the changes" +
                  "When there is silence, freeze";

//Frequency-to-Tap-Dance Practice

<<(sine)______~~~~~~~~~___/\___~~~~~~>> [pause: 500ms]
<<(sine)___\_/~~~\/~___\_/\_/~~~\____________>> [pause: 1000ms]
<<(sine)____/~~~~~>> [pause: 500ms]
<<(sine)____/~~~~~____/~~~~~>> [pause: 2000ms]
<<(sine)__/~~~\_/\/~~~~~~\____/\/\/\/\/\\\\////~~~~~~~~~>>
```

Audio Objects and Image Objects are characterized by inherent differences when communicating PerPL instructions. Audio Objects are finite, time-dependent representations requiring subsequent Audio Objects to be concatenated or scheduled. Image Objects are instantaneous, infinitely static representations requiring subsequent Image Objects to be manually triggered or scheduled. Interrupted Audio Objects risk obscuration when instruction information is incomplete. Image Objects always render complete information and risk obscuration when the human processing the image is interrupted.

Audio instruction timing depends on the length of the audio instruction relative to how the audio communicates meaning: words, sounds, contour, transients, associations, music or patterns. Audio instruction length and efficiency must perpetually balance with the time required to perform the instruction. Rapid performance changes require succinct audio instructions to indicate the changes as fast as they are performed. Triggering new audio instructions before performers have completed previous instructions shifts the balance in the other direction. Harnessing tighter isomorphic relationships between the process-describing characteristics of audio instructions and the processual qualities driving performatizations will enhance timing stability.

Image Object timing is determined by human interpretation capacity relative to the image's function in PerPL. An Image Object of music notation may require each note to be slowly performed. The same image of music notation can come and go in a flash provoking a fleeting notion of music before boots, arrows, and popsicles are seen. An Image Object of only red may cue a sudden gesture before a green image is displayed cueing a different gesture. The same red image can function as an indefinite temporal marker for performing a deep introspective consideration of the color red.

### Performatized Branching

The appropriate strategy for programming a conditional statement in PerPL relies on the nature of its impelementation. Imagine 50 performers either sitting or standing. All of the performers receive the same audio instructions from Telebrain, an implementation SIMD processing. The PerPL programmer wishes to send the following conditional statement:

```
if (standing) sit;
else stand;
```

The simplest solution is to save the above pseudocode as a Text-To-Speech Audio Object named `Stand/Sit-Switch`. Since the pseudocode is understandable in everyday language, the instruction needs no translation. When `Stand/Sit-Switch` is distributed during a performance, all performers hear the same spoken text instructing the standing performers to sit and the sitting performers to stand.

Sending `Stand/Sit-Switch` several times in a row toggles the performers between states of sitting and standing, regardless of their initial state. If a PerPL programmer plans to use an instruction repeatedly during a performance, the programmer can assign the instruction to a sound, as follows:

```
<<BLEEEEP>> =  if (standing) sit;
               else stand;
```

The programmer creates the Uploaded Audio Object `Bleeeep` by uploading the sound `<<BLEEEEP>>` to Telebrain. The text, "When you hear this sound," is saved as a Text-To-Speech Audio Object named `When`. To build an assignment training instruction, the programmer concatenates several Audio Objects into an Audio Sentence named `BleeeepTraining.`

```
BleeeepTraining  =  When + Bleeeep + Sit/Stand-Switch;
```

Prior to using `Stand/Sit-Switch` in a Performance, the programmer sends the `BleeeepTraining` instruction. Later, when the programmer wishes a `Stand/Sit-Switch` to be performed, only `Bleeeep` is sent. Assigning functions to sounds contributes to timing precision by clarifying when instructions are performed and allowing subsequent instructions to follow in quicker succession.

If "sit" and "stand" are used in conjunction with "spin once," "spin twice," "squat while raising your right arm" and "touch your toes," then an alternate programming strategy is preferred. With training, performers can learn to associate the six actions with six sounds. According to research in sonification and auralization, associating data, or an action, with the semiotic meaning of a sound contributes to learning and retention.[8]

To implement the new strategy, the PerPL programmer uploads six sounds into Audio Objects and saves descriptions of the six actions into Text-To-Speech Audio Objects. The pseudocode representing the uploaded sounds are delimited by double angle brackets. The texts within the double angle brackets are onomatopoeic representations of how the Audio Objects sound, illustrating semiotic associations with the actions they represent.

```
//6-Actions Training

6-ActionsTraining = "When you hear each sound" +
                    "perform the action that follows" +
                    <<SHLOOEEP>> + "stand" +
                    <<PLEEOOSH>> + "sit" +
                    <<SHWISH>> + "spin once" +
                    <<SHWISHWISH>> + "spin twice" +
                    <<BRAAMFDING>> + "squat while raising your right arm" +
                    <<WEWAWOWU>> + "touch your toes";

//6-Actions Practice

"While each sound plays, perform its action"  [pause: 3000ms]
<<SHLOOEEP>> [pause: 1000ms]
<<PLEEOOSH>> [pause: 1000ms]
<<SHWISH>> [pause: 1000ms]
<<SHWISHWISH>> [pause: 1000ms]
<<BRAAMFDING>> [pause: 600ms]
<<WEWAWOWU>> [pause: 600ms]
<<BRAAMFDING>> [pause: 600ms]
<<SHWISHWISH>> [pause: 3000ms]
<<PLEEOOSH>> [pause: 2000ms]
<<SHLOOEEP>> [pause: 5000ms]
<<SHWISH>> [pause: 500ms]
<<WEWAWOWU>>
```

To integrate `6-Actions` with `Stand/Sit-Switch`, additional sounds are associated with `if` and `else`. The programmer assigns `if` to a 1000Hz sine tone. To implement an `if` statement, the sound of an action is understood as a comparison because it is overlaid with a 1000Hz sine tone — the sound of the action under evaluation and the 1000Hz sine tone are heard at the same time. To implement Stand/Sit-Switch, <<SHLOOEEPP>> will represent both the state of *standing* and the instruction to *stand*. To test *if* a performer is *standing,* the sound <<SHLOOEEPP>> is played simultaneously with a 1000Hz sine tone. The next sound played represents the action to perform in the first branch of the conditional statement.

To implement the else branch of the conditional statement, else is assigned to the sound <<CLUNK>>. Since else does not require a comparison, the sound is played unlayered. The sound that follows <<CLUNK>> represents the action to perform in the second branch of the conditional statement.

PerPL programmers can layer audio by saving Audio Objects into an Audio Layer Object. In the example below, the pseudocode LAYER [ AudioObject1, AudioObject2 ] is used to represent an Audio Layer Object.

```
//If-Else Training

If-ElseTraining = "if you hear the sound for an action" +
                  "overlaid with this sound" + <<1000HZ SINE TONE>> +
                  "ask yourself if you are performing the action" +
                  "if you are" +
                  "perform the action for the next sound you hear" +
                  "if you are not" +
                  "do not perform the next sound you hear" +
                  "when you hear the sound" + <<CLUNK>> +
                  "if you did not perform the previous sound" +
                  "perform the action for the next sound you hear";

//If-Else-Practice

LAYER [ <<1000HZ SINE TONE>>, <<SHLOOEEP>> ]
<<PLEEOOSH>>
<<CLUNK>>
<<SHLOOEEP>>

LAYER [ <<1000HZ SINE TONE>>, <<PLEEOOSH>> ]
<<SHLOOEEP>>
<<CLUNK>>
<<PLEEOOSH>>
```

```
LAYER [ <<1000HZ SINE TONE>>, <<SHLOOEEP>> ]
<<SHWISH>>
<<CLUNK>>
<<BRAAMFDING>>

LAYER [ <<1000HZ SINE TONE>>, <<BRAAMFDING>> ]  // if(a){b} if(c){a} else{d}
<<WEWAWOWU>>
LAYER [ <<1000HZ SINE TONE>>, <<SHWISH>> ]
<<BRAAMFDING>>
<<CLUNK>>
<<SHWISHWISH>>

LAYER [ <<1000HZ SINE TONE>>, <<BRAAMFDING>> ]
<<SHLOOEEP>>
<<CLUNK>>
<<PLEEOOSH>>
```

Limiting performatizations to conditionals and actions can generate complex results. `If-Else-Practice` illustrates the use of PerPL instructions made entirely from non-speech sounds. As PerPL instructions become more complex, PerPL intepreters and their training routines develop as well.

At which point will PerPL audio instructions be repurposed as music?

# THE FUTURE

The future of Telebrain is divided into immediate goals of increased functionality and distant glimpses of what could become.

### Immediate Future

The immediate goals of Telebrain extend current functionality, stimulate PerPL development, and incorporate potent research from other fields. Future Telebrain developments are listed as follows:

- **Add sound synthesis -** A browser-based sound synthesis engine reduces the latency of serving audio files in real-time. Greater varieties of modifiable sound will augment meaning transmission potential. Currently on hold until robust, cross-platform libraries are available.
- **Implement spearcons**[9] - Spearcons are time-compressed words or phrases that are played too fast to be recognized. Developed for auralization, spearcons outperform regular speech in experimental data. Since they are lexical, they exploit the language processing centers of the brain. Currently on hold until a free, variable-speed, cross-platform text-to-speech solution is available. The current text-to-speech implementation resolves compatibility issues that arise with the variable-speed alternatives.
- **Utilize built-in client device features** - Most computers and mobile devices include GPS, accelerometers, built-in microphones, built-in cameras, text-to-speech utilities and the ability to vibrate. Future implementation depends on cross-platform browser access to these features.
- **Improve usability and explainability** - To maximize potential uses and users, emphasize user-centered development by making clear and simple interfaces for complex processes. Advocate Telebrain's use across disciplines.
- **Simulate evolutionary dynamics in PerPL's development architecture** - Rebuild the the PerPL instruction programming architecture to capitalize on existing evolutionary dynamics of human social systems. Investigate evolutionary algorithms appropriate for network-based collective language development and model the processes into Telebrain's PerPL development architecture.

- **Automate PerPL instruction generation** - Implement user-definable Telebrain templates for automatically generating generic performatization schemes. Performance architectures defined by network topologies of programmer/performer relationships can be preset. Using commonly implemented PerPL syntax and semantic rules to automatically generate PerPL instructions will simplify the programming process — limiting configurability when desired. Templates invite new PerPL programmers to experience Telebrain with less investment. Advanced PerPL programmers can develop templates for reusable performance architectures and PerPL rules.
- **Develop interfaces for isomorphic transference** - Implement sonification and auralization techniques for generating PerPL instructions. Develop interfaces for easy translations between datasets and algorithms into PerPL instructions.
- **Create a text-based version of PerPL** - Design a Telebrain text editor for a specified, extensible text-based PerPL. Text-based PerPL statements and codeblocks will be automatically translated into multi-media PerPL instructions. Incorporate schemes for converting text-based PerPL into the grammar of spoken language, and vice-versa.
- **Integrate models of human cognition** - Model the neurobiological timing thresholds of the human listening process into parameters of PerPL instructions. Focusing, aiming, or distributing heirarchies of algorithmic processes into specified temporal ranges of aural perception will enhance the interpretability of audio instructions.
- **Incorporate principles of ABL (A Behavior Language)** - ABL is an artificial behavioral programming language used to control digital autonomous agents in virtual interactive environments.[10] Making artificial interactions seem real, behavioral programming languages are primarily used in virtual environments and video games. Applying principles of ABL to real-time, real-space, real-human performances may reveal new techniques for organizing performance instructions. ABL supports joint action by systematically coordinating the behavior of multiple performers.[11] Future actions of performers depend on the results of current performer actions. The following is an example of ABL code implemented when there is a knock at the door:

```
sequential behavior AnswerTheDoor() {
    WME w;

    with success_test { w = (KnockWME) } wait; act sigh();
    subgoal OpenDoor();
    subgoal GreetGuest();
    mental_act { deleteWME(w); }
}

sequential behavior OpenDoor() {
    precondition { (KnockWME doorID :: door)
          (PosWME spriteID == door pos :: doorPos)
          (PosWME spriteID == me pos :: myPos)
          (Util.computeDistance(doorPos, myPos) > 100)
    }
    specificity 2;
    // Too far to walk, yell for knocker to come in subgoal
    YellAndWaitForGuestToEnter(doorID);
}

sequential behavior OpenDoor() {
    precondition { (KnockWME doorID :: door) }
    specificity 1;
    // Default behavior - walk to door and open
}
```
[12]

**Distant Future**

Continuous integration of heterogeneous configurations fuels the locomotion towards the boundary of imaginations. Communication is the medium — where signals signal signals. Communication as computation, literally conceived as metaphor, is elaborated into the future using vibrational computation.

Historically, proof of artificial intelligence is thought to manifest in conversations between human and machine.[13] When the conversation is unsatisfactory, half of the conversation is physically reconfigured to dictate a different conversation.

Human symbolic language co-evolved with paralinguistic prosodic capabilities. Equivalent paralinguistic functions are rarely found in artificial symbolic languages such as mathematics and computer programming languages. In contrast, the artificial symbolic languages used to notate music represent parameters of sound that correlate with paralinguistic characteristics. Musical expression, often changes in volume, rhythm and pitch, can provoke internal experiences considered difficult to describe with words.

The space between language and music is an area of underutilized capacity where current modes of communication can expand. Due to the tight evolutionarily entanglement between our hyper-sensitive oral-aural capacity, cognitive infrastructure and language facility, sound is a logical nursery for growing fresh branches of communication. Since sound is carried by waves, a future computational paradigm is imagined as intersecting vibrations.

Vibrational computation occurs when waves co-compute midstream. The intersection of computing waves outputs interference pattern results. A multiplexed wave embodying program, data, interpreter, and self-describing co-signal-processing ability intersects with a similarly data-enriched wave. Always containing incomplete information, each signal relies on interaction to fill in the gaps. As the waves intersect, the combined instructions interlock. Through superimposed micro-interference, canceling and reinforcing as they perpetually go, the multiplexed waves co-interpret each other.

The evolution of symbolic language distanced humans from their subjective experiences. Once experiences were untethered from experiencer, they could be interpreted, shared and

reconfigured — recursively filtered through generations of brains.

Vibrational computation places interpretation in the space between communicators, creating actual interpretive distance. The external interpreter is conjured by the act of communication — as if each signal contained the genetic information for creating an instantaneous brain at every shared intersection. Vibrational computation resonates through self-describing wave-driven zippers interlocking patterns like gamelans through space.

# CONCLUSION

Telebrain is a platform for entangling systems of meaning, a place where paradoxes self-resolve through the evolutionary potential introduced by incorporating human interaction into the computational process.

Computers evolved due to generations of humans externalizing their internal cognitive structures.[14] Holding up a labyrinthian mirror, Telebrain externalizes the computer's internal cognitive structures into novel paradigms of human performance. Communication is the common thread, where incongruous systems of logic, language and art commingle, configure, and transform.

# Appendix I. Telebrain Functionality ( http://telebrain.org )

## CONTENT OBJECTS

### Audio Objects

Web-Based Audio: Web-based audio allows online audio files to be available during a performance. Copyrighted material is not allowed. When saving Web-Based Audio Telebrain makes a local MP3 copy of the audio on the Telebrain sever. Currently, only audio file types supported by the user's browser are supported.

Uploaded Audio: Audio uploads copy audio content from a client device to the server. Only MP3 audio files are supported at this time.

Text-To-Speech Audio**:** Text-To-Speech audio can be saved in advance or generated in real-time during a performance. To make Text-To-Speech audio in advance, choose a language and then save up to 100 characters of text. When the save button is pressed, Telebrain saves an mp3 of the Text-To-Speech audio to the server. Text-To-Speech audio can be accessed during a performance or concatenated with other audio using the Audio Sentence functionality.

Audio Collections**:** Since Audio Sentences and Audio Layers generate a new audio files when saved, these Collections function as Audio Objects. Audio Sentence and Audio Layers can be incorporated into new Collections, however altering the original Audio Objects after they are used in Audio Sentences or Audio Layers does not ensure that preexisting Collections will be updated. Generally, new audio files are generated at the time the Collections are saved to Telebrain - re-saving the referencing Collections updates the audio files to the current version. This may be automated in future versions of Telebrain.

### Image Objects

Web-Based Images: Web-based images allow graphic internet content to be made visible during a performance. To save a web-based image, find a valid URL linking directly to online image content. A valid image URL with end with '.jpg' or '.png', and should not be a link to an html page containing the image. The easiest way to obtain a valid image URL is to right-click on the desired image in order to open the image in a new tab or window

Uploaded Images: Image Uploads allow content uploaded from a computer or mobile device to be available during a performance. The upload functionality is currently suspended. In the meantime, email the image to Telebrain and the image will be uploaded manually.

Teleprompts: (Text-To-Image) Teleprompts graphically display text during a performance. Parameters such as font, size, text color, and background color can be assigned to each Teleprompt Object.

## COLLECTION OBJECTS

**Folder Objects:** Folders are unordered collections of Telebrain Content and Programs. Folders are an organizational tool and allow associated data to be assigned to particular Roles during a performance. Folders can be used to filter content used in a particular performance and can also be used in Timed Organization Algorithms.

**Audio-Image Pairs:** Audio-Image Pairs allow Image Objects and Audio Objects to be distributed simultaneously to a Role during a performance. Audio-Image Pairs function similarly to other Telebrain Content and automatically adjust to Roles with limited Audio / Image receiving functionality.

**Audio Sentence Objects:** Audio Sentences are ordered collections of Audio Objects that are concatenated into a single audio file. Audio Sentence create a new concatenated audio file in order to reducing the timing inconsistencies of delivering multiple smaller audio files individually. The start times of individual audio files in a generated Audio Sentence can be used to quickly change the order the Audio Sentence is played after the audio is delivered to the performers. When a new Audio Sentence is saved, the Collection can be used as an Audio Object.

**Audio Layer Objects:** Audio Layers allow Audio Objects to be played simultaneously during a performance. Each Audio Object layer requires an assigned start-time and relative volume level. The Audio Objects in an Audio Layer are mixed and saved to a new audio file. When new Audio Layers are saved, the Collection can be used as an Audio Object.

**Image Phrase Objects:** Image Phrases are ordered Collections of Image Objects available for use during a performance. Image Phrases can be stepped through manually during a performance or used in Organization Algorithms.

## PROGRAMS

### Program Setup

Role Objects: Roles organize possible performance functions allowing each performer or performer group to have their send/receive capabilities and interface layout assigned in advance. Multiple performers can play the same role in a performance. The maximum number of performers per Role can be limited in the Venue model. Checkboxes turn the following visibility and functionality on or off for each Role: Telebrain Menu, Performance Title, Send Text, Send Text-To-Speech (live), Send Image, Send Audio, Send Association, Send Fraction, Send OSC, Receive Text, Receive Text-To-Speech (live), Receive Image, Receive Audio, Receive Interface, Receive OSC, Role List, Performer List, Performer Activity Log, Global Activity Log, Change Role, Change Interface, Change Functionality, Test Functionality. See Performance Functionality for more information

Venue Objects: Venues are models of performance architecture. The parameters assigned to each venue will determine how Roles interact in a performance. When performing, a new instance of the venue model is loaded allowing performers to join by selecting a Role with predetermined functionality. Venue models may need to be declared before creating Associations, Algorithms, and Content assignments depending on the interrelational requirements of a particular performance. Venue declarations can limit the number of performers allowed per Role. The type of information needed when joining a performance is determined such as requiring a nickname, local IP address, or passcode.

Interface Objects: Interfaces allow Content, Collections, Associations, and Algorithms to be assigned to user interface elements such as buttons, pull-down menus, text inputs, and display areas. Interfaces can be dynamically rendered, associated with Roles, and/or incorporated into Collections, Associations, and Algorithms.

### Assignments

Multi-Role Assignments: Multi-Role Assignments allow Content to be associated with multiple Roles in advance. Multi-Role Assignments can distribute a variety of Content simultaneously to different Roles during a live performance. Multi-Role Assignments can be used in Timed Organization Algorithms in order to construct other multi-performer distribution presets. Multi-Role Assignments are dependent on a pre-existing venue because the available roles must be known in advance. Multi-Role Assignments are also dependent on the functionalities assigned to each Role, because the associated roles must be able to receive the assigned content.

Fractional Assignments: Fractional Assignments associate different Content with fractions of a group or Role (unlike Multi-Role Assignments where different content is assigned to different roles). Fractional Assignments default to having ALL performers as the group to be divided, but the assignments can also be associated to a particular Role or Roles. Fractional Assignments have two modes of dividing a performer group. Persistent Mode remembers the Fractions performers are assigned to throughout the performance. Dynamic Mode randomly divides performers into new Fractions each time the Fractional Assignment is called.

### Algorithms

Timers: Timers of particular lengths of time can be saved in advanced for use during a performance. Timers are necessary for synchronizing events on multiple devices due to latency inconsistencies encountered delivering data from the server to multiple clients. Telebrain timers are self-adjusting and continually synchronize to the server clock in order to assure time-accuracy. Although immediately synchronized events cannot be guaranteed, events can be triggered simultaneously after a predetermined amount of delay.

Metronomes: Metronomes allow multiple performance events to be triggered at a regular timed interval. Multiple Metronomes can be synchronized to the server or run at varying tempi without synchronization.

OSC Objects: OSC (Open Sound Control) Objects allow incoming and outgoing OSC address parameters to be associated with Telebrain Content, Collections, and Programs. OSC Objects allow Telebrain to interface with external OSC-capable hardware and software, and control information to be routed to clients through a local network bypassing the remote Telebrain server.

Timed Organization: Timed Organization Algorithms allow Timers and Metronomes to be assigned to OSC Objects, Content, Collections, Roles, Venues, Interfaces, and Associations. This is where timing functions can be assigned to algorithmically organize all other Telebrain functionality.

Distribution Organization: Distribution Organization Algorithms allow OSC Objects, Content, Collections, Roles, Venues, Interfaces, and Associations to algorithmically organized without timed restriction.

## PERFORMANCE

**Start Performance:** To start a new Performance, the first performer must instantiate a Venue model by selecting from a list of Venues and naming the Performance. Once the first performer has chosen a Role and provided a nickname and/or IP address, if required, the Performance exists and becomes available for additional performers to join. The Performance can be protected by an associated passcode allowing only performers with the correct passcode to join.

**Join Performance:** If a Performance exists, a “Join Performance” pull-down menu listing current Performances is rendered allowing performers to select the Performance to join. The performer selects a Role and if required, enters a unique nickname and/or their local IP address.

### Performance Functionality

Telebrain Menu: Show or Hide the Telebrain Navigation menu. Hiding creates a larger display area for Image Content and Interface, but limits Telebrain navigation during a Performance.

Performance Title: Show or Hide the name of the current Performance. Hiding allows more display area for Image Content and Interface, but multi-Venue Performances may require performers to be aware of their current Performance name.

Send Text: Show or Hide text input interface for typing and sending real-time text during a Performance. Send Text functions as a chatroom text input and can be routed to some or all performers depending on routing assignments and the receiving Roles’ functionality.

Send Text-To-Speech (live): Show or hide text input interface for typing and generating real-time Text-To-Speech Audio during a Performance. The Send Text-To-Speech functions as a chatroom text input, except the received text is received as Text-To-Speech Audio. The Text-To-Speech audio can be routed to some or all performers depending on routing assignments and the receiving Roles’ functionality.

Send Image: Show or hide a pull-down menu listing the Image Content available to a particular Venue and/or Role. If no Image Content has been assigned, then all Telebrain Image Content is listed. When selected, the Image Content is immediately sent to some or all performers depending on routing assignments and the receiving Roles’ functionality.

Send Audio: Show or hide a pull-down menu listing the Audio Content available to a particular Venue and/or Role. If no Audio Content has been assigned, then all Telebrain Audio Content is listed. When selected, the Audio Content is immediately sent to some or all performers depending on routing assignments and the receiving Roles’ functionality.

Send Multi-Role: Show or hide a pull-down menu listing the Audio Content available to a particular Venue and/or Role. If no Audio Content has been assigned, then all Telebrain Audio Content is listed. When selected, the Audio Content is immediately sent to some or all performers depending on routing assignments and the receiving Roles’ functionality.

Send Fraction: Show or hide a pull-down menu listing the Fractional Assignments available to a particular Venue and/or Role. If no Fractional Assignments have been associated, then all Telebrain Fractional Assignments appropriate to the Venue model will be listed. When selected, the Content associated to each Fraction is immediately sent to the performers divided into Fractions. Fractional Assignments send Audio and/or Image Content to associated performers.

Send OSC: Show or hide a pull-down menu listing the OSC Presets available to a particular Venue and/or Role. When selected, the preset OSC message is immediately sent to the local IP addresses associated with some or all performers depending on routing assignments and the receiving Roles’ functionality.

Send Algorithm: Show or hide a pull-down menu listing the Algorithms available to a particular Venue and/or Role. If no Algorithms have been associated, then all Telebrain Algorithms appropriate to the Venue model will be listed. Depending on the selected Algorithm, new Interface elements may appear depending on the input requirements of the Algorithm. Control of the Algorithm will be made available to the sender.

Receive Text: Allow received text to be displayed.

Receive Text-To-Speech (live): Allow the received audio file generated from real-time Text-To-Speech text to be played on the local device.

Receive Image: Allow received Image Content to be displayed.

Receive Audio: Allow received Audio Content to be played on the local device.

Receive Interface: Allow received Interface Content to be displayed and used.

Receive OSC: Allow received OSC messages to trigger other Telebrain functionality.

Role List: Show or Hide the list of Roles associated with the Venue model of the Performance. Checkboxes are displayed next to each Role in the list to indicate the Roles to which Content will be routed. Associations, Fractions, and Algorithms override Role Routing Assignments.

Performer List: Show or Hide the nicknames of current performers and indicate their Role. Checkboxes are displayed next to each Nickname in the list to indicate the specific Performers to which Content will be routed. Associations, Fractions, and Algorithms override Performer Routing Assignments.

Performer Activity Log: Show or Hide a list of the times all Instructions Sent or Received by the Performer .

Global Activity Log: Show or Hide a list describing all Instructions Sent or Received by all Performers.

Change Role: Show or Hide a pull-down menu of Roles associated with the Venue. The performer switches to the selected Role.

Change Interface: Show or Hide a pull-down menu of Interfaces associated with the Role and/or Venue. The display shows or hides the selected Interface.

Change Functionality: Show or Hide a pull-down menu of Performance Functionalities, allowing the current Performance Functionalities to changed during a Performance.

Test Functionality: Show or Hide the necessary interface elements for Testing currently assigned Functionalities. When Testing, the performers can send themselves Content to test their receive settings or receive Content to test their send settings.Test functionalities can be controlled by a master prompter in the form of a Sound Check Algorithm or Image Check Algorithm.

Leave Performance: A link that reads “X Leave Performance” is displayed in the top-left corner of all live Performances. Clicking on this link removes the performer from the current Performance and returns to the Telebrain website.

Audio Required: If a Role indicates Audio Receive functionality is required, a link that reads “Audio Required” is displayed to the right of the “X Leave Performance” link. Clicking the “Audio Required” link turns Audio On if the audio is turned off and turns Audio Off if the audio is already on. The Speaker in the upper-right corner of the Telebrain navigation menu indicates the current Audio state as well, but may not be visible if Telebrain Menu is hidden. When Audio is switched from off to on, the “Telebrain” audio file is played.

## GENERAL FUNCTIONALITY

**Speaker Icon:** The icon in the top-right corner of the Telebrain Navigation Menu is red when Audio is Off and green when Audio is On. When Audio is switched from Off to On, the “Telebrain” audio file is played. The Speaker Icon must be green indicating that the Audio is On for Audio Content to be played by Telebrain.

**Lock/Unlock:** Content, Collections, and Programs can be locked and unlocked in order to protect user-generated data. A passcode is required when locking saved data and can only be unlocked and therefore edited when the same passcode is entered. Since all data stored on Telebrain can used in any Performance, locking and unlocking data protects the information from being altered or deleted by another user.

**Acknowledgments**

This work was supported by the Herb Alpert School of Music at the California Institute of the Arts and benefited from the valuable input of my doctoral committee: David Rosenboom, Sara Roberts, and Robert Wannamaker.

I would like to thank Mark Trayle, Tim Perkis, Clay Chaplin, Anne LeBaron, Lauren Pratt, Andrea Young, Bill Gribble, Hugh Robert MacMillan, STEIM, WACM and the Improvised Music Theater classes at California Institute of the Arts for additional ideas and support.